\documentclass[12pt]{iopart}

\usepackage{amssymb}
\usepackage{graphicx}

\begin{document}

\article[Investigating the EGS region at LHC with polarized and
unpolarized final states]{Heavy ion collisions at the LHC, last call
  for predictions}{\center Investigating the extended geometric
  scaling region at LHC with polarized and unpolarized final states}

\author{Dani\"el Boer, Andre Utermann, \underline{Erik Wessels}}

\address{Department of Physics and Astronomy,
VU University Amsterdam, \\
De Boelelaan 1081, 1081 HV Amsterdam, The Netherlands}

\eads{\mailto{D.Boer@few.vu.nl}, \mailto{A.Utermann@few.vu.nl},
\mailto{E.Wessels@few.vu.nl}}

\begin{abstract}
We present predictions for charged hadron production and $\Lambda$
polarization in $p\,$-$p$ and $p\,$-$Pb$ collisions at the LHC using
the saturation inspired DHJ model for the dipole cross section in the
extended geometric scaling region.
\end{abstract}

%\maketitle

%%%%%%%%%%%%%%%%%%%%%%%%%%%%%%%%%%%%%%%%%%%%%%%%%%%%%%%%%%%%%%%%%%%%%%%%%%

%\section*{Introduction}

At high energy, scattering of a particle off a nucleus can
be described in terms of a colour dipole scattering off small-$x$
partons, predominantly gluons, in the nucleus. At very high energy
(small $x$), the dipole amplitude starts to evolve nonlinearly with
$x$, leading to saturation of the density of these small-$x$ gluons.
The scale associated with this nonlinearity, the saturation scale
$Q_s(x)$, grows exponentially with $\log(1/x)$.

The nonlinear evolution of the dipole amplitude is expected to be
characterized by geometric scaling, which means that the dipole
amplitude depends only on the combination $r_t^2Q^2_s(x)$, instead of
on $r_t^2$ and $x$ independently. Moreover, the scaling behaviour is
expected to hold approximately in the so-called extended geometric
scaling (EGS) region between $Q_s^2(x)$ and $Q^2_{gs}(x)\sim
Q_s^2(x)/\Lambda$.

The small-$x$ DIS data from HERA, which show geometric scaling, were
successfully described by the GBW model \cite{Golec-Biernat:1998js}.
To describe the RHIC data on hadron production in $d\,$-$Au$ in the
EGS region a modification of the GBW model was proposed by Dumitru,
Hayashigaki and Jalilian-Marian (DHJ), incorporating scaling
violations in terms of a function $\gamma$\footnote{We note that at
  central rapidities we cannot reproduce exactly the results of
  \cite{Dumitru:2005kb} for large $p_t$. Therefore, a modification of
  the model may be needed to describe all RHIC
  data.} \cite{Dumitru:2005kb}. This DHJ model also describes $p\,$-$p$
data at forward rapidities \cite{Boer:2006rj}.

\section*{DHJ model prediction for charged hadron production}

Using the DHJ model we can make a prediction for the $p_t$-spectrum of
charged hadron production in both $p\,$-$Pb$ and $p\,$-$p$ collisions
at the LHC, at respectively $\sqrt{s}=8.8$ TeV and $\sqrt{s}=14$ TeV.
Figure \ref{fig}a shows the minimum bias invariant yield for an
observed hadron rapidity of $y_h=2$ in the centre of mass frame, which
for $1$ GeV $\lesssim p_t\lesssim 10$ GeV predominantly probes the EGS
region.  We note that at this rapidity the result is not sensitive to
details of the DHJ model in the saturation region $r_t^2>1/Q^2_s$.
Further, from \cite{Dumitru:2005kb} we expect that $p_t$-independent
$K$-factors are needed to fix the normalization. We conclude that the
LHC data on hadron production in both $p\,$-$Pb$ and $p\,$-$p$
collisions will provide valuable data to further study the dipole
scattering amplitude near the onset of saturation, particularly the
behaviour of the function $\gamma$, which is discussed in e.g.\
\cite{Boer:2007wf}.

\section*{DHJ model prediction for $\Lambda$ polarization}

\begin{figure}[htb]
\includegraphics*[height=53mm]{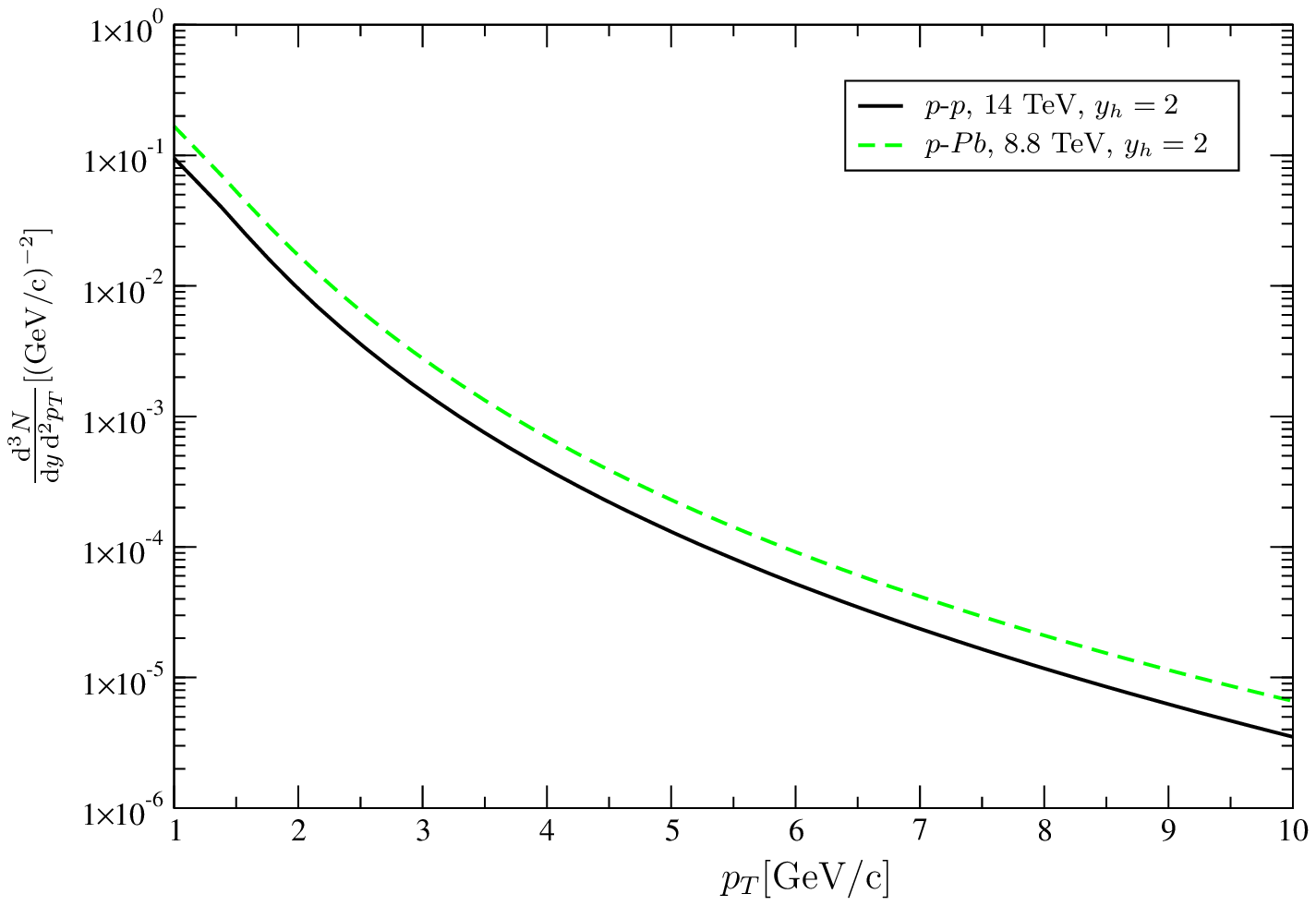}
\includegraphics*[height=53mm]{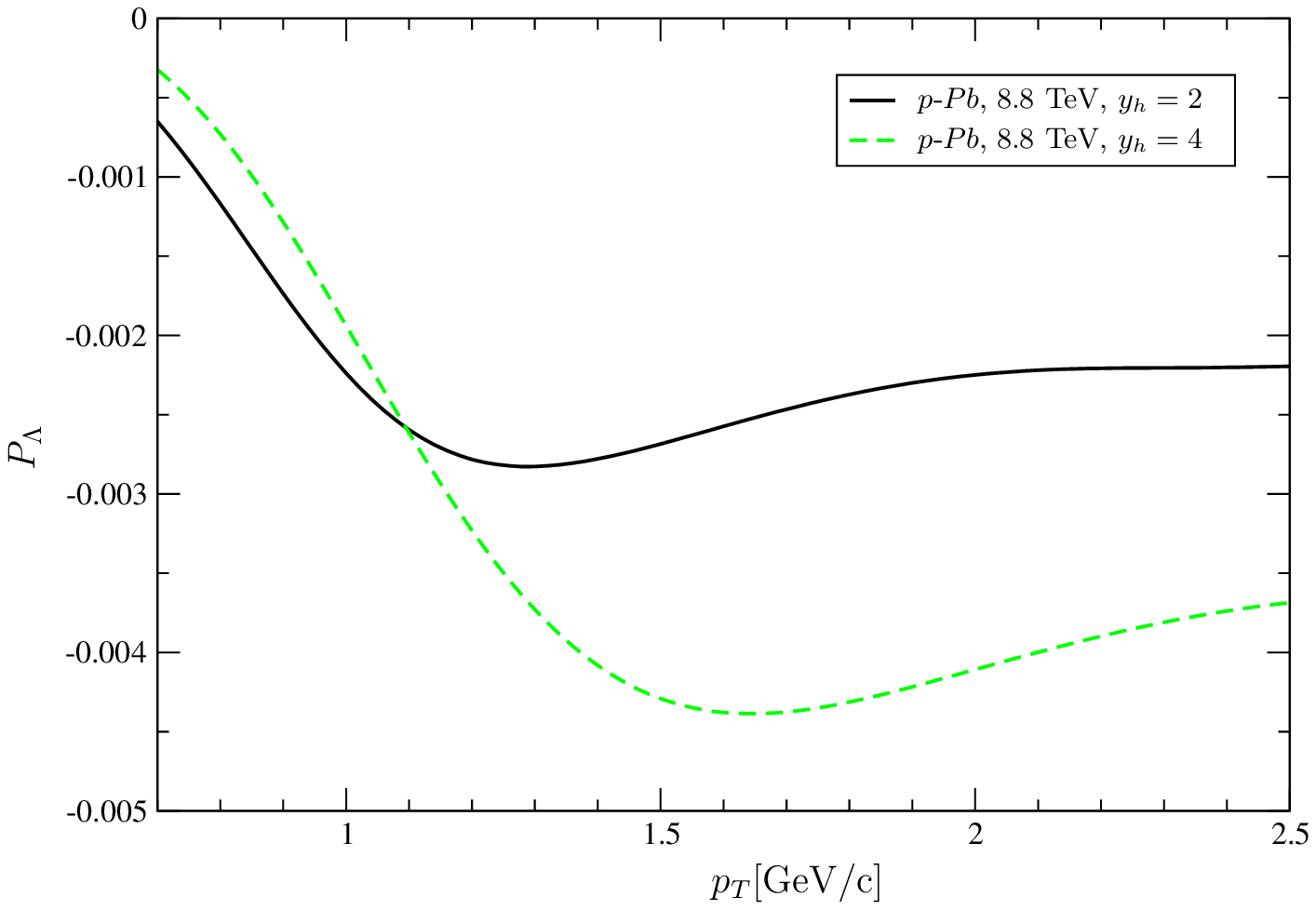}
\caption{\label{fig}a.\ Charged hadron production. b.\ $\Lambda$
  polarization. In both plots, $A_{\rm eff}=20$, and parton
  distributions and fragmentation functions of \cite{Dumitru:2005kb}
  and \cite{Boer:2002ij} were used.}
\end{figure}

Another interesting small-$x$ observable is the polarization of
$\Lambda$ hyperons produced in $p\,$-$A$ collisions, $P_\Lambda$. This
polarization, oriented transversely to the production plane, was shown
to essentially probe the derivative of the dipole scattering
amplitude, hence displaying a peak around $Q_s$ when described in the
McLerran-Venugopalan model \cite{Boer:2002ij}.  If this feature
persists when $x$-evolution of the dipole scattering amplitude is
taken into account, $P_\Lambda$ would be a valuable probe of
saturation effects. Using the DHJ model for the $x$-evolution of the
scattering amplitude, we find that $P_\Lambda$ displays similar
behaviour as in the MV model. This is depicted for fixed $\Lambda$
rapidities of 2 and 4 in figure \ref{fig}b. The position of the peak
scales with the average value of the saturation scale $\langle
Q_s(x)\rangle$. In the plotted region, the peak is located roughly at
$\langle Q_s(x)\rangle/2$.

The figure also shows that, like in the MV model, in the DHJ model
$|P_\Lambda|$ scales approximately linearly with $x_F$, which means
that at the LHC it is very small due to $\sqrt{s}$ being very large:
rapidities around 6 are required for $P_\Lambda$ to be on the $1\%$
level, although there is a considerable model uncertainty in the
normalization.

We conclude that the polarization of $\Lambda$ particles in $p\,$-$Pb$
collisions is an interesting probe of $\langle Q_s(x)\rangle$, but is
probably of measurable size only at very forward rapidities.

~\\[-1mm]
We thank Adrian Dumitru and Jamal Jalilian-Marian for helpful
discussions.

\end{document}